\documentclass[twocolumn,preprintnumbers,amsmath,amssymb]{revtex4}
\usepackage{graphicx}
\usepackage{graphicx,ulem,amsmath,amssymb,amsfonts,xcolor}

\usepackage{bm}			
\bibpunct{}{}{,}{s}{}{} 

\newcommand\beq{\begin{equation}}
	\newcommand\eeq{\end{equation}}

\begin{document}
	\title{Terahertz generation via all-optical quantum control in 2D and 3D materials}
	
	\author{Kamalesh Jana$^1$} \email{kjana@uottawa.ca}
	\author{Amanda B. B. de Souza$^1$, Yonghao Mi$^1$, Shima Gholam-Mirzaei$^1$, Dong Hyuk Ko$^1$, Saroj R. Tripathi$^1,^2$, Shawn Sederberg$^3$, James A. Gupta$^4$}
	\author{Paul B. Corkum$^1$}  \email{pcorkum@uottawa.ca}
	
	\affiliation{$^1$Joint Attosecond Science Laboratory, University of Ottawa and National Research Council Canada, 25 Templeton Street, Ottawa, ON K1N 6N5, Canada}
	\affiliation{$^2$Graduate School of Science and Technology, Shizuoka University, 3-5-1 Johoku, Hamamatsu, Shizuoka, 432-8561, Japan}
	\affiliation{$^3$School of Engineering Science, Simon Fraser University, 8888 University Drive, Burnaby, BC, V5A 1S6, Canada}
	\affiliation{$^4$Department of Physics, University of Ottawa, Ottawa, ON K1N 6N5, Canada}
	
	\begin{abstract}
		{Using optical technology for current injection and electromagnetic emission simplifies the comparison between materials.  Here, we inject current into monolayer graphene and bulk gallium arsenide (GaAs) using two-color quantum interference and detect the emitted electric field by electro-optic sampling.  We find the amplitude of emitted terahertz (THz) radiation scales in the same way for both materials even though they differ in dimension, band gap, atomic composition, symmetry and lattice structure. In addition, we observe the same mapping of the current direction to the light characteristics. With no electrodes for injection or detection, our approach will allow electron scattering timescales to be directly measured. We envisage that it will enable exploration of new materials suitable for generating terahertz magnetic fields.}
	\end{abstract}

	\maketitle
	
Isolated terahertz magnetic fields are hard to produce, yet they have unrivaled potential for controlling magnetic materials on an ultrafast time scale.\cite{KirilyukRMP2010,YamaguchiPRL2010,YamaguchiPRL2013} One approach to generating intense magnetic pulses has been to use the magnetic field that accompanies the current of the high energy electron beam.\cite{TudosaNat2004,BackSci1999}  While effective, this approach is complex, inflexible, and inefficient. A more practical approach is to adapt THz technology,\cite{KochNatRev2023,LeeTHzBook2009} designed for linearly polarized radiation, to generate azimuthal polarization with spatially varying polarization, where the electric field vector aligns along the azimuthal direction (around the propagation axis).\cite{ZhanAOP2009} In this structured light pulse, the magnetic field is isolated at the center of the THz beam.\cite{HellwarthPRE1996,ZdagkasNatPhot2022,JanaSciAdv2024}  Any THz technology that allows an electric field of $10^{6}$ V/cm or more for linearly polarized THz beams should be adaptable to produce an isolated magnetic field at the center of a focused azimuthally polarized beam of $\sim$ 0.33 Tesla or more. Solid-state materials like semiconductors and semi-metals are technologically relevant platforms for generating THz magnetic fields.\cite{JanaSciAdv2024,SederbergNatPhot2020,JanaNatPhot2021,JanaNanophot2022,SederbergAPL2022}

Quantum control proves remarkably efficient in injecting directional currents by leveraging quantum interference between absorption pathways induced by two harmonically-related laser pulses.\cite{DupontPRL1995,ShapiroBook2012,AtanasovPRL1996,CostaNatPhys2007,SederbergPRX2020} In this process the current density ($J$) injection rate is given by the relation\cite{JanaNatPhot2021,DupontPRL1995,ShapiroBook2012,AtanasovPRL1996,CostaNatPhys2007}  $\frac{dJ}{dt} = 2\eta E_\omega E_\omega E_{2\omega} sin(\Delta\phi)$. Here, $\eta$ is a constant related to fourth rank tensor element, $E_\omega$ and $E_{2\omega}$ are field amplitudes of fundamental ($\omega$) and second harmonic (2$\omega$)  pulses respectively and $\Delta\phi$ denotes relative phase between two pulses. Using this approach, we have demonstrated significant current injection into GaAs.\cite{JanaSciAdv2024,SederbergNatPhot2020,JanaNatPhot2021,JanaNanophot2022,SederbergAPL2022} A key feature of GaAs has been that we can generate any structured current and hence any structured THz beam consistent with Maxwell’s equations, using only circularly polarized fundamental, linearly polarized second harmonic and a spatial light modulator to control their relative phase in a pixel by pixel manner.\cite{JanaNatPhot2021,JanaNanophot2022,SederbergAPL2022} Such control enables the generation and reconfiguration of space-time-coupled toroidal terahertz pulses and isolated magnetic fields.\cite{JanaSciAdv2024}
 
Despite the flexibility in THz waveform and spatial structure enabled by quantum controlled currents, scaling the THz power radiated from bulk semiconductors is limited by several factors. Phase walk-off between the two-color pulses, saturation effects, and attenuation of the THz radiation by the free electron plasma, all contribute to this limit.\cite{CoteAPL1999,DarrowIEEEQE1992,TaniApplOptics1997} Additionally, the shorter scattering time of the current limits the frequency content and amplitude of the emitted single-cycle THz pulse (see Supplemental Material, Fig. S1\cite{SupplementalMaterial}).

In this context, graphene\cite{GeimNatMat2007,NetoRMP2009}, a carbon-based monolayer semi-metal with zero-bandgap energy represents an ideal platform for quantum control. Moreover, as a 2-D material, scattering in graphene is restricted to two dimensions with an additional contribution from material-substrate interaction. With a high optical nonlinearity, high damage threshold, and longer relaxation timescale, several layers of graphene having spacing between layers could potentially enable scaling to higher THz amplitudes than what is possible in bulk semiconductors such as GaAs.

Theoretical studies have shown that current injection is possible in monolayer graphene \cite{MelePRB2000,RiouxPRB2011,RaoPRB2012} through quantum interference between single- and two-photon absorption, similar to effects observed in bulk semiconductors. While all-optical current injection and measurement have been demonstrated in multilayer epitaxial graphene\cite{SunNanoLett2010} and graphite\cite{NewsonNanoLett2008}, such noncontact measurements in monolayer graphene have not yet been reported. Recently, currents have been injected and controlled in monolayer graphene using near single-cycle laser pulse.\cite{HiguchiNat2017,HeidePRL2018,HeideNanophot2021} However, all experiments thus far have employed an integrated electrode geometry for current measurement\cite{HiguchiNat2017,HeidePRL2018,HeideNanophot2021}, making it difficult to distinguish between electronic processes occurring in the graphene and those happening at the graphene-electrode interface.

Here, we demonstrate two-color quantum control in chemical vapor deposition (CVD) grown monolayer graphene on sapphire by measuring the emitted terahertz radiation. We quantitatively compare the results from graphene, a zero-bandgap semi-metal, having a symmetric planar structure and composed of only carbon with GaAs, a traditional semiconducting electro-optical material with a bandgap of 1.43 eV and lacking inversion symmetry. Surprisingly, aside from a much smaller amplitude for graphene, we only see a small difference in the THz signal. Both scale with intensity in the same way. Initially we use linearly polarized control-pulses for simplicity. Using a combination of circularly polarized fundamental and linearly polarized second harmonic pulses, we also show 360$^{\circ}$ control over current direction in graphene. If, as is proposed, petahertz electronics becomes available from graphene, quantum control will be the method by which we inject real or virtual currents and we will monitor using direct emission.

	\begin{figure}
	
	\includegraphics[width=1.0\columnwidth]{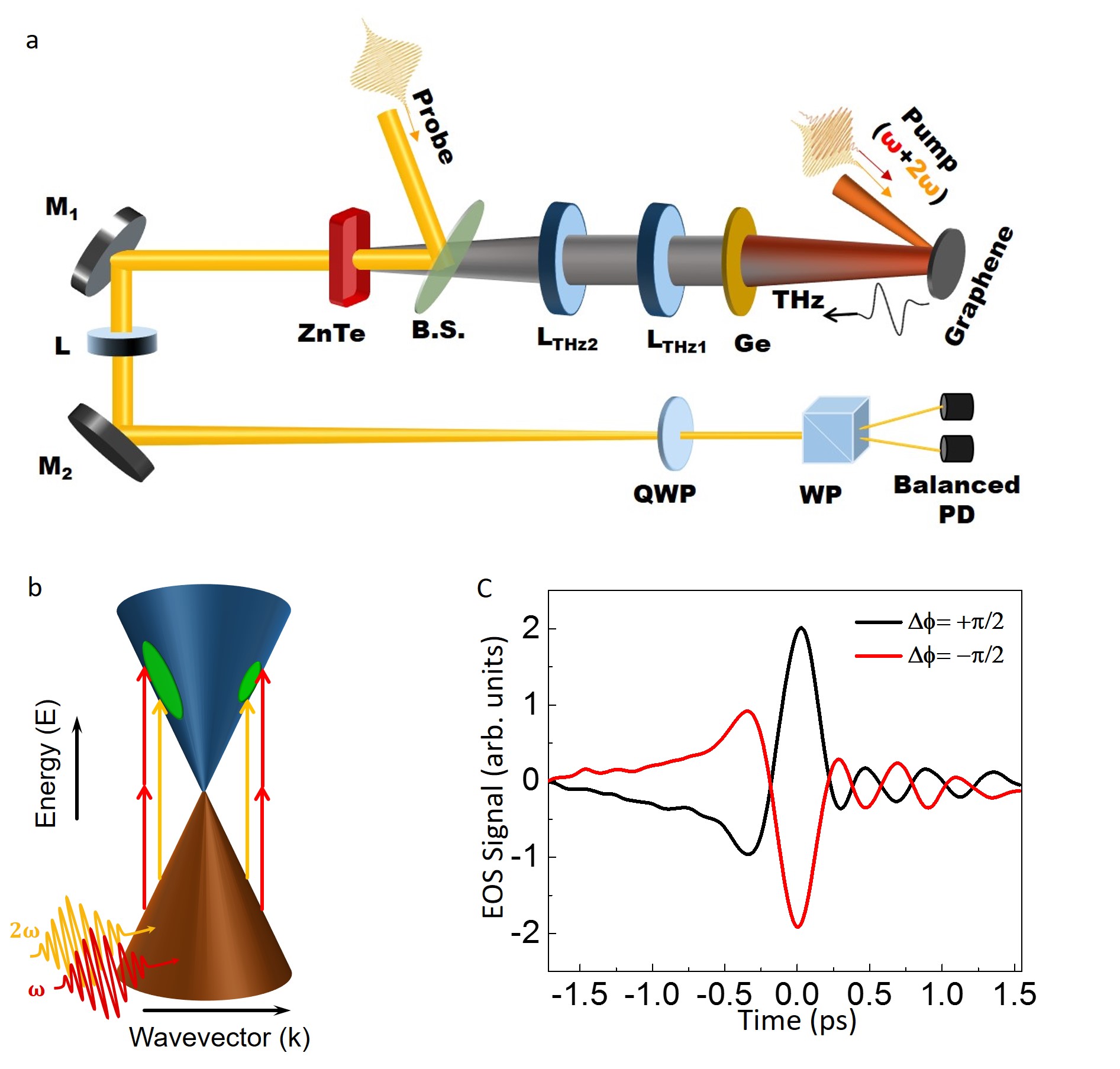}
	\caption{ Generation and measurement of THz pulse under excitation of two-color laser fields. (a) Experimental setup for measuring THz radiation emitted from ultrafast injected currents in single-layer graphene. Two-color laser pulse (1480 nm + 740 nm) generates currents in graphene that radiates THz pulse. Ge: germanium wafer, $M_1-M_2$: metallic mirrors, L: focusing lens, $L_{THZ1}-L_{THZ2}$: THz lenses, QWP: quarter-wave plate, WP: Wollaston prism, BS: pellicle beam splitter, PD: photodiode, ZnTe: zinc telluride crystal. (b) Illustration shows simultaneous single and two-photon electronic transitions in graphene, resulting asymmetric charge distribution. (c) Measured THz pulses from monolayer graphene. Time dependent electro-optic signals (EOS) for $\Delta\phi=+\pi/2$ (black) and $\Delta\phi=-\pi/2$ (red) are presented.}
	\label{figure1}
\end{figure}

We use femtosecond two-color laser pulses ($\lambda_\omega$=1480 nm and $\lambda_{2\omega}$=740 nm) for THz generation [Fig. 1a] in monolayer graphene or bulk GaAs wafer (see Supplemental Material, Note 2\cite{SupplementalMaterial}). The optical parametric amplifier (OPA), operating at a 1 kHz repetition rate, generates a 40 fs fundamental laser pulse at 1480 nm. A $\beta$-barium borate (BBO) crystal is used to generate a second-harmonic pulse at 740 nm. The two beams are combined and focused to a 700 $\mu$m diameter spot on the sample. The two-color laser fields induce quantum interference between single- and two-photon transitions [Fig. 1b], resulting in a phase-controlled transient photocurrent at the sample, that radiates a single cycle THz pulse. We employ free space electro-optic sampling\cite{QiAPL1995} (EOS) to characterize the radiated THz pulse [Fig. 1a]. We collect the radiated THz beam from the front side of the sample in a reflection geometry using a THz lens (f/1.5). The second THz lens (f/2) is used to refocus the THz onto 1 mm thick zinc telluride (ZnTe) crystal wherein THz pulse is sampled. We use a 0.5 mm thick germanium substrate as a low-pass filter that blocks the residual $\omega$ and 2$\omega$ pump beam and transmit THz. A small part of the second harmonic (740 nm) beam is used to sample the THz pulse.

The temporal traces of the radiated THz field from graphene under excitation of two co-polarized laser fields are shown in Fig. 1c. Here, traces are presented for two different values of relative phase ($\Delta\phi$) of the two-color fields. It clearly depicts that the polarity of the THz waveform flips after changing the relative phase ($\Delta\phi$) by $\pi$. A more detailed dependence of the radiated THz field on the relative phase ($\Delta\phi$) (see Supplemental Material, Fig. S3\cite{SupplementalMaterial}) shows that the peak THz electric field changes its amplitude and sign as the relative phase ($\Delta\phi$) is adjusted. Sinusoidal oscillation clearly indicates that the radiated THz field is generated from quantum controlled current source induced by two-color laser fields. Notably, the dependence of THz amplitude and polarization on the relative phase offers a remarkable level of control.

\begin{figure}
	
	\includegraphics[width=1.0\columnwidth]{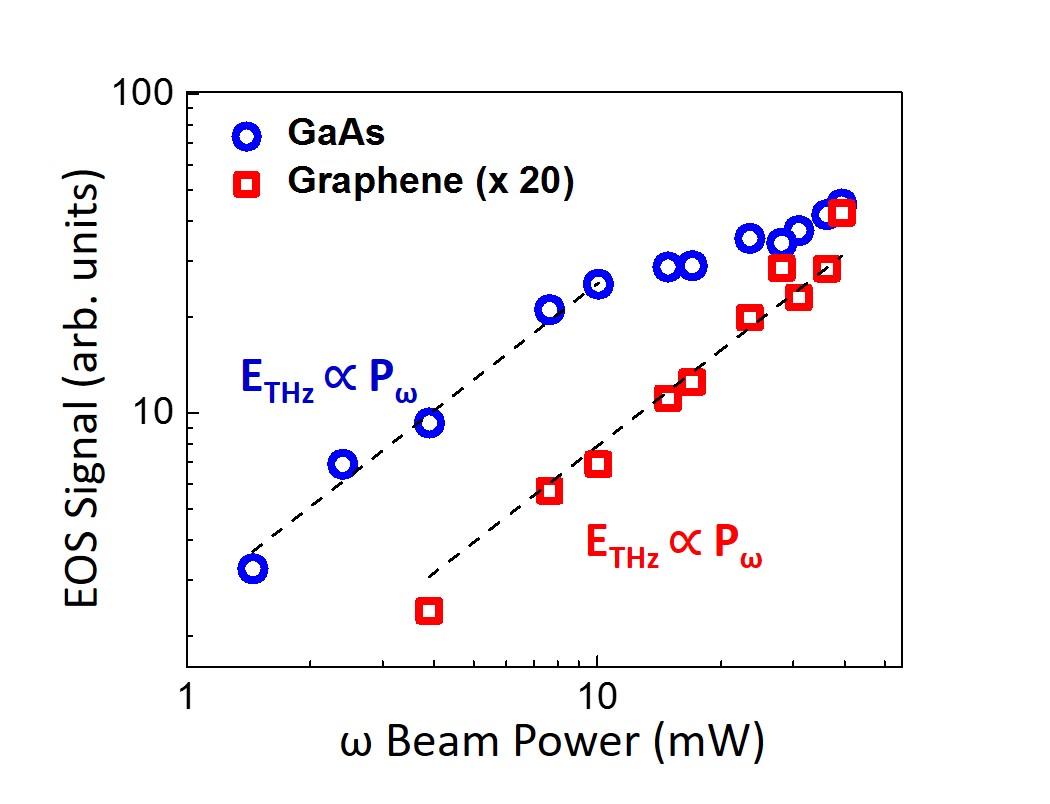}
	\caption{ Dependence of emitted THz field on pump power. Peak THz fields (or EOS Signals) from single-layer graphene and 110 $\mu$m thick GaAs samples, observed as a function of the average power of the incident fundamental pulse ($\omega$) when power of the second harmonic pulse (2$\omega$) is fixed. To represent in the same plot THz peak fields from graphene are multiplied by a constant factor 20.}
	\label{figure2}
\end{figure}

Figure 2 presents the dependence of the THz field on the average power of the fundamental ($P_{\omega}$) pulse while the power of the second harmonic pulse ($P_{2\omega}$) is kept constant. We measure the power dependence of THz emission for monolayer graphene and 110 $\mu$m thick bulk GaAs samples. For the monolayer graphene, we observe a linear increase in the THz field ($E_{THz}$) with the fundamental pump power ($P_{\omega}$), as expected from a quantum-controlled process, i.e., $E_{THz} \propto E_{\omega} E_\omega E_{2\omega} sin(\Delta\phi) \propto P_{\omega} \sqrt{P_{2\omega}} sin(\Delta\phi)$. However, for the bulk GaAs sample, we observe that the peak THz field initially increases linearly with the $\omega$ beam power and after 10 mW the slope starts saturating. The saturation of THz field at high pump power may be attributed to screening effects and attenuation of THz radiation by dense free electron plasma.\cite{CoteAPL1999,DarrowIEEEQE1992,TaniApplOptics1997}
 
We also see that at the highest applied pump power, the peak THz field from the graphene is approximately 20 times smaller than that obtained from the GaAs sample using the same laser parameters. However, it is important to note that in GaAs, the quantum interference takes place within a layer $\sim$ 500 nm thick below the surface, limited by the penetration depth\cite{EngNanophot2015} of the second harmonic pulse (2$\omega$). Hence, the THz signal in bulk GaAs originates from multiple atomic layers whereas, in the case of graphene, only a sub-nanometer thick single layer generates THz radiation. From this perspective, it is possible that graphene is more efficient at generating THz radiation compared to its semiconductor counterparts. We expect that using multiple layers of graphene, each spaced with h-BN, will significantly enhance THz radiation efficiency, potentially surpassing that of GaAs. 

\begin{figure}
	\includegraphics[width=1.0\columnwidth]{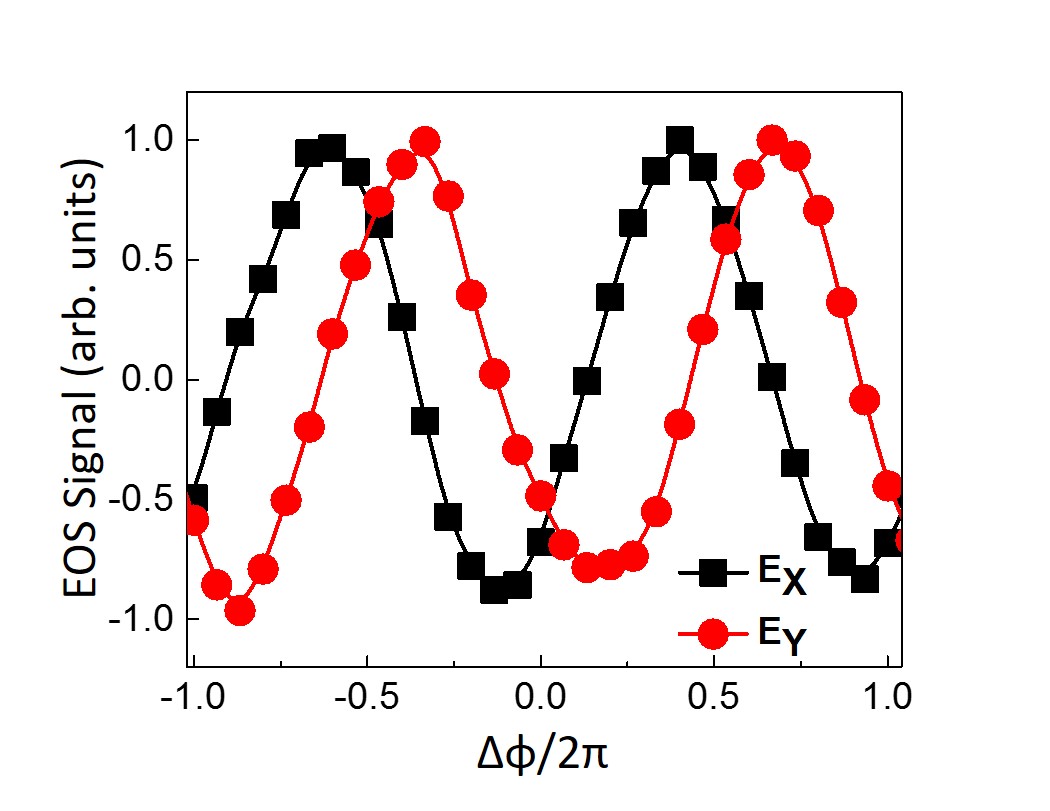}
	\caption{ Polarization control of the THz pulse emitted from graphene. A combination of circularly polarized $\omega$ and linearly polarized 2$\omega$  fields is used to control the polarization of the radiated THz pulse. x and y-components ($E_X$ and $E_Y$) of the radiated THz field are measured as a function of relative phase ($\Delta\phi$) of the two-color laser pulse. Both components ($E_X$ and $E_Y$) are measured at the peak of the THz waveforms. }
	\label{figure3}
\end{figure}

To understand the influence of beam polarization on the radiated THz from graphene, we first conduct measurements with linearly polarized pump pulses. We used a THz wire grid polarizer to measure the polarization of the radiated THz pulse. We observe that the THz radiation exhibits almost the same polarization as the two co-polarized pump beams (see Supplemental Material, Fig. S4\cite{SupplementalMaterial}).
Next, we used a combination of circularly polarized fundamental ($\omega$) beam and linearly polarized second-harmonic beam (2$\omega$) and measure the radiated THz from graphene. We measure the x- or y-components ($E_X$ and $E_Y$) of the THz field as a function of relative phase ($\Delta\phi$) between two laser fields.  Figure 3 shows that the two field components are oscillating $\pi$/2 out of phase as we adjust the relative phase ($\Delta\phi$). We plot $E_X$ and $E_Y$ parametrically as a function of $\Delta\phi$ over one cycle (see Supplemental Material, Fig. S5\cite{SupplementalMaterial}). The figure is almost a circle, a field of constant amplitude, which demonstrates that by changing the relative phase the polarization axis of the radiated THz pulse rotates and with it the direction of the injected current.

 The same relationship between the current direction and light characteristics was observed in GaAs\cite{JanaNatPhot2021}. Orientation manipulation of THz electric field could find many applications in imaging and spectroscopy. For example, linear dichroism and birefringence\cite{JordensApplOpt2009,HiroyaOptMatExp2023} of wide variety of materials can be measured using terahertz time domain spectroscopy without rotating the sample under investigation.

Finally, in Fig. 4a we compare the frequency spectrum of the THz signal (see Supplemental Material, Fig. S6\cite{SupplementalMaterial}), as measured through our optical system, from GaAs and graphene.  We see there is a small difference.  The GaAs spectrum is slightly blue shifted compared to the graphene spectrum.  We associate this shift to the lower density of states available in graphene to electrons leading to a lower scattering rate.

 Since the current is injected fast (in our case 40 fs, but this could be decreased to 10 fs) we can have a source in the high frequency region while the low frequency content of the THz signal will be a direct measurement of the current decay (or the electron scattering rate). The former opens the possibility of producing very high-power flying doughnut pulses at 30 THz\cite{JanaSciAdv2024,CorkumIEEEQElc1985} while the latter opens the potential to study electron scattering in different materials.
 
  We numerically calculate the THz waveforms by varying current relaxation times (see Supplemental Material\cite{SupplementalMaterial}). Our results indicate that, in addition to changes in amplitude, the peak frequency of the THz pulse shifts with relaxation time (see Supplemental Material, Fig. S1\cite{SupplementalMaterial}). Figure 4b presents the normalized THz spectra for four different current relaxation timescales. We see as the current relaxation time increases, the peak of the THz spectrum shifts towards lower frequencies. It is important to mention that in the experiment; we were unable to capture the entire bandwidth of the THz pulse due to the very weak electro-optic response of the detection crystal (ZnTe) beyond 5 THz. Additionally, numerical aperture of the collection optics limits the collection efficiency of the low frequency part of the THz pulse. However, based on our measurements, we predict that the current relaxation timescale in monolayer graphene will be longer than that in GaAs.\cite{LeitensdorferPRL1996}

\begin{figure}
	
	\includegraphics[width=1.0\columnwidth]{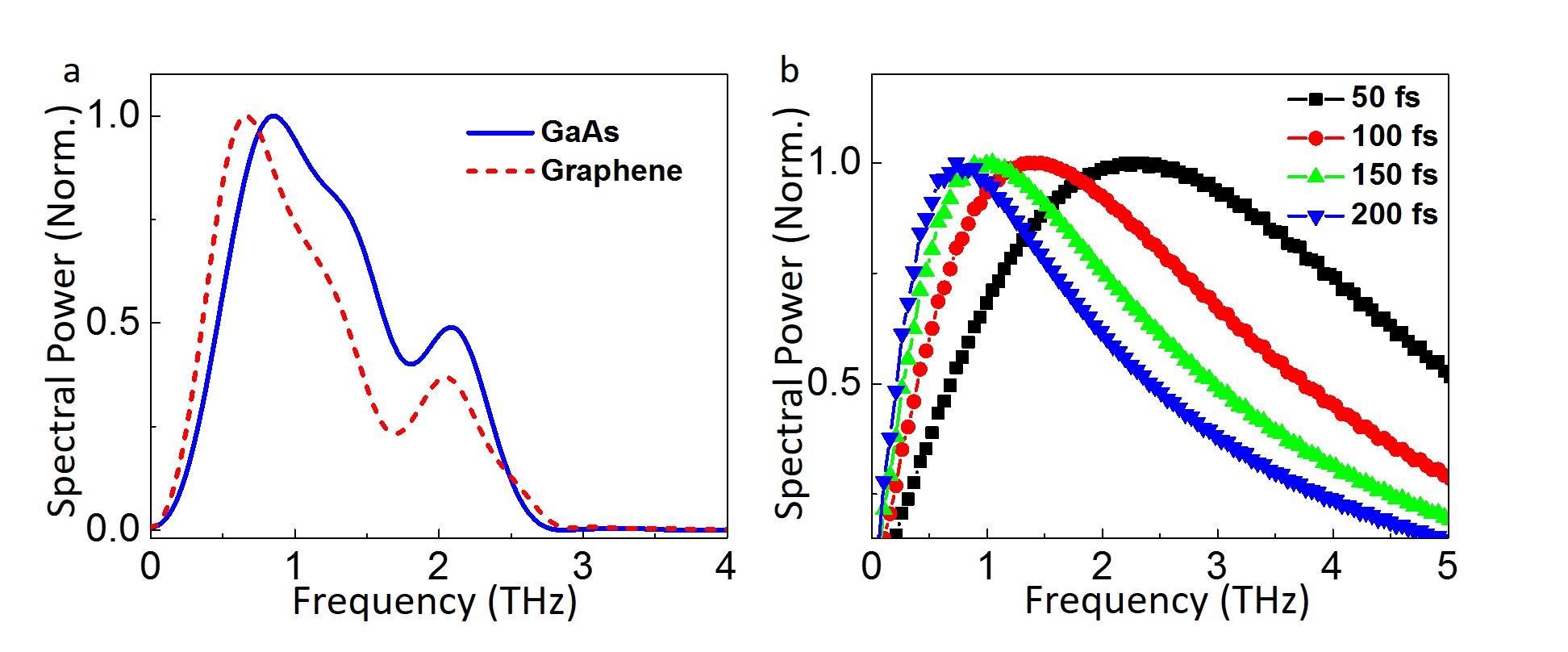}
	\caption{THz emission spectra. (a) Frequency spectrum of THz pulses measured from single-layer graphene (red dashed) and bulk GaAs (blue solid) samples under the same experimental condition. Our measurements cannot capture the full spectrum of the THz pulses due to the limitations of the collection optics and the frequency response of the detection crystal. (b) Calculated THz spectra for four different current relaxation times (50 fs, 100 fs, 150 fs, 200 fs). THz Spectra are normalized for better representation.}
	\label{figure4}
\end{figure}


In conclusion, we have measured THz pulses radiated from a monolayer graphene irradiated by a two-color femtosecond laser pulse. We observe that the amplitude and polarity of the THz pulse is controllable with the relative phase of two-color pulse, confirming that the current injection results from quantum interference. We have systematically investigated the effect of pump polarization and pump power scaling on the radiated THz pulse. We show that using a circularly polarized pump pulse enables the rotation of the polarization of the radiated THz pulse, thereby controlling the direction of the injected current. These results establish graphene as a promising platform for generating structured currents, space-time-coupled THz pulses, and isolated magnetic fields. Our all-optical current injection and measurement approach also provides some insight into carrier scattering timescales.

Our measurements show that although the radiated THz field from the monolayer graphene is much weaker than that measured from a bulk semiconductor (GaAs), considering the thickness and absorption lengths of both samples, graphene may prove to be an efficient source of THz radiation. We used an undoped graphene sample, a zero band gap semimetal with the Fermi level located at the junction of the two conical bands, enabling single-photon absorption at both pump wavelengths ($\omega$ and 2$\omega$). This may decrease the available fundamental beam ($\omega$) power for two-photon absorptions. It is important to note that, in graphene photoelectrons generated from single-photon absorption of fundamental beam ($\omega$) do not contribute to the quantum interference process and hence the background plasma density, electron-hole scattering and free carrier absorption will increase. Optimizing the Fermi level in graphene samples through doping or applying gate voltage might offer significant advantages. Additionally, it may be important to consider using a semiconducting 2-D material instead of graphene. 

We have presented the generation and measurement of THz radiation in the far field, created from a monolayer graphene and we quantitatively compare it with radiation generated in bulk GaAs, both pumped with $\lambda_\omega$=1.48 $\mu$m light and its second harmonic. While we use a single layer of graphene directly grown on sapphire, graphene can be grown on other material\cite{ScarfeScRep2021} allowing us to tune and minimize the scattering time. We expect that using multilayer graphene, each layer spaced with h-BN, or 2-D semiconducting substrates will enhance the THz radiation efficiency significantly.  Furthermore, in the same lab we are constructing a picosecond, 200 mJ laser (Yb: YAG) pulse that can be used as a pump, so we can increase the incident power by a factor of nearly $\sim10^{3}$.   Finally, optimizing the experiments with, for example, better collection optics will also improve the overall THz field (with better telescope design, we have already improved our peak B-field by a factor of 5). Even with these changes, if the graphene sample is used in reflection, we can make a quantitative comparison with GaAs. We expect that employing two-color vector laser pulses will generate a high-density ring current\cite{JanaSciAdv2024}, capable of producing Tesla-scale isolated THz magnetic field pulses from 2-D solid-state materials, joining near-field multi-Tesla scale pulses generated from gases.\cite{SederbergPRX2020} Together, these sources will enable us to control magnetic materials in picosecond times.\\

The authors are grateful for important discussions with Jean-Michel Menard. This research was supported by the Natural Sciences and Engineering Research Council of Canada (NSERC) Discovery Grant Program, University of Ottawa Distinguished Faculty program, and the United States Army Research Office (award number: W911NF-19-1-0211). A.B.B.D. acknowledges the support of the Undergraduate Research Opportunity Program (UROP) from the University of Ottawa.


\begin{thebibliography}{}
		
\bibitem{KirilyukRMP2010} A. Kirilyuk, A. V. Kimel and T. Rasing, Rev. Mod. Phys. \textbf{822,} 2731 (2010).

\bibitem{YamaguchiPRL2010}	K. Yamaguchi, M. Nakajima and  T. Suemoto, Phys. Rev. Lett. \textbf{105,} 237201 (2010).

\bibitem{YamaguchiPRL2013}	K. Yamaguchi, T. Kurihara, Y. Minami, M. Nakajima, T. Suemoto, Phys. Rev. Lett. \textbf{110,} 137204 (2013).

\bibitem{TudosaNat2004}	I. Tudosa, C. Stamm, A. B. Kashuba, F King , H. C. Siegmann, J. Stöhr, G. Ju, B. Lu,  B. Weller, Nature \textbf{428,} 831 (2004).

\bibitem{BackSci1999}	C. H. Back, R. Allenspach, W. Weber, S. S. P. Parkin, D. Weller, E. L. Garwin, and H. C. Siegmann, Science  \textbf{285,} 864  (1999).

\bibitem{KochNatRev2023}  M. Koch, D. M. Mittleman, J. Ornik, E. Castro-Camus, Nat. Rev. Methods Primers \textbf{3,} 48 (2023).

\bibitem{LeeTHzBook2009} Y. S. 	Lee, \textit{Principles of Terahertz Science and Technology} (Springer, New York, USA, 2009).

\bibitem{ZhanAOP2009} Q. Zhan, Adv. Opt. Photonics \textbf{1}, 1 (2009).

\bibitem{HellwarthPRE1996} R. W. Hellwarth, P. Nouchi, Phys. Rev. E \textbf{54,} 889 (1996).

\bibitem{ZdagkasNatPhot2022} A. Zdagkas,  C. McDonnell, J. Deng, Y. Shen, G. Li, T. Ellenbogen, N. Papasimakis,  N.I. Zheludev, Nat. Photonics\textbf{16,} 523 (2022).

\bibitem{JanaSciAdv2024} K. Jana, Y. Mi, S. H. Møller, D. H. Ko, S. Gholam-Mirzaei, D. Abdollahpour, S. Sederberg, P. B. Corkum, Science Advances \textbf{10,} eadl1803 (2024).

\bibitem{SederbergNatPhot2020}	S. Sederberg, F. Kong, F. Hufnagel, C. Zhang,E. Karimi, P.B. Corkum, Nat. Photonics \textbf{14,} 680 (2020). 

\bibitem{JanaNatPhot2021} K. Jana, K. R.  Herperger, F. Kong, Y. Mi, C. Zhang, P.B. Corkum, S. Sederberg, Nat. Photonics \textbf{15,} 622  (2021). 

\bibitem{JanaNanophot2022}	K. Jana, E. Okocha, S. H. Møller, Y. Mi, S. Sederberg, P.B. Corkum, Nanophotonics \textbf{11,} 787 (2022).

\bibitem{SederbergAPL2022}	S. Sederberg, P. B. Corkum,  Appl. Phys. Lett. \textbf{120,} 160504 (2022).

\bibitem{DupontPRL1995}	E. Dupont, P. B. Corkum, H. C. Liu, M. Buchanan,  Z. R. Wasilewksi, Phys. Rev. Lett. \textbf{74,} 3596 (1995).

\bibitem{ShapiroBook2012}	M. Shapiro, P. Brumer, \textit{Quantum control of molecular processes} (John Wiley Sons, 2012).

\bibitem{AtanasovPRL1996}	R. Atanasov, A. Haché, J. L. P. Hughes, H. M. van Driel, J. E. Sipe, Phys. Rev. Lett. \textbf{76,} 1703 (1996).

\bibitem{CostaNatPhys2007} L. Costa, M. Betz, M. Spasenovic, A. D. Bristow, H. M.  van Driel, Nature Phys. \textbf{3,} 632 (2007).

\bibitem{SederbergPRX2020} S. Sederberg, F. Kong, P.B. Corkum,  Phys. Rev. X \textbf{10,} 011063 (2020).

\bibitem{CoteAPL1999} D.  Côté, J. M. Fraser, M. DeCamp, P. H. Bucksbaum, H. M. van Driel, Appl. Phys. Lett. \textbf{75,}  3959 (1999).

\bibitem{DarrowIEEEQE1992} J. T. Darrow, X. C. Zhang, D. H. Auston, J. D. Morse, IEEE J. Quantum Electron. \textbf{28,}  1607 (1992).

\bibitem{TaniApplOptics1997} M. Tani, S. Matsuura, K. Sakai, S. I. Nakashima, Applied Optics \textbf{36,}  7853 (1997).

\bibitem{SupplementalMaterial}See Supplemental Material

\bibitem{GeimNatMat2007} A. K. Geim, K. S. Novoselov, Nat. Mater.\textbf{6,}  183 (2007).

\bibitem{NetoRMP2009} A. H. C. Neto, F. Guinea, N. M. R. Peres, K. S. Novoselov, A. K. Geim, Rev. Mod. Phys.\textbf{81,}  109 (2009).

\bibitem{MelePRB2000} E. J. Mele, P. Král, D. Tománek, Phys. Rev. B \textbf{61,}  7669 (2000).


\bibitem{RiouxPRB2011}J. Rioux, G. Burkard, and J. E. Sipe, Phys. Rev. B \textbf{83,} 195406 (2011).

\bibitem{RaoPRB2012}K. M. Rao and J. E. Sipe, Phys. Rev. B \textbf{86,} 115427 (2012).

\bibitem{SunNanoLett2010} D. Sun, C. Divin, J. Rioux,, J. E. Sipe,  C. Berger, W. A. Heer,    P. N. First, and T. B. Norris,  Nano Lett. \textbf{10,} 1293–1296 (2010).

\bibitem{NewsonNanoLett2008} R. W. Newson, J. M. Ménard, C. Sames, M. Betz, H. M. van Driel, Nano Lett. \textbf{8,} 1586-1589 (2008).

\bibitem{HiguchiNat2017} T. Higuchi, C. Heide, K. Ullmann, H. B. Weber, P. Hommelhoff, Nature \textbf{550,}  224 (2017).


\bibitem{HeidePRL2018} C. Heide, T. Higuchi, H. B. Weber, P. Hommelhoff, Phys. Rev. Lett. \textbf{121,}   207401, (2018).

\bibitem{HeideNanophot2021} C. 	Heide, T. Boolakee, T. Eckstein, P. Hommelhoff, Nanophotonics \textbf{10,} 3701 (2021).

\bibitem{QiAPL1995} W. Qi, X‐C. Zhang, Appl. Phys. Lett. \textbf{67,} 3523 (1995).

\bibitem{EngNanophot2015} P. C. Eng, S. Song, B. Ping, Nanophotonics \textbf{4,} 277 (2015).

\bibitem{JordensApplOpt2009} C. Jördens, M. Scheller, M. Wichmann, M. Mikulics, K. Wiesauer, M. Koch, Applied optics \textbf{48,} 2037 (2009).

\bibitem{HiroyaOptMatExp2023} 	I. Hiroya, K. Takeya, S.R. Tripathi, Optical Materials Express \textbf{13,} 966 (2023).

\bibitem{CorkumIEEEQElc1985} P. B. Corkum, IEEE J. Quant. Elect. \textbf{21,} 216 (1985).


\bibitem{LeitensdorferPRL1996} A. Leitensdorfer, C. Fürst, A. Lauberau, W. Kaiser, G. Tränkle, and G. Weimann, Phys. Rev. Lett. \textbf{76,} 1545 (1996).

\bibitem{ScarfeScRep2021} S. Scarfe, W. Cui, A. Luican-Mayer, J.-M. Menard, Sci. Rep. \textbf{11,} 8729 (2021). 

 
 

		
		
		
		
	\end{thebibliography}
\end{document}